# A simple, distance-dependent formulation of the Watts-Strogatz model for directed and undirected small-world networks

H. Francis Song[1] and Xiao-Jing Wang[1, 2]

[1]*Center for Neural Science, New York University, New York, NY 10003*
[2]*NYU-ECNU Joint Institute of Brain and Cognitive Science, NYU Shanghai, Shanghai, China*
(Dated: August 22, 2014)

Small-world networks—complex networks characterized by a combination of high clustering and short path lengths—are widely studied using the paradigmatic model of Watts and Strogatz (WS). Although the WS model is already quite minimal and intuitive, we describe an alternative formulation of the WS model in terms of a distance-dependent probability of connection that further simplifies, both practically and theoretically, the generation of directed and undirected WS-type small-world networks. In addition to highlighting an essential feature of the WS model that has previously been overlooked, this alternative formulation makes it possible to derive exact expressions for quantities such as the degree and motif distributions and global clustering coefficient for both directed and undirected networks in terms of model parameters.

PACS numbers: Complex Systems, Biological Physics, Interdisciplinary Physics

Many biological, technological, and social networks have the 'small-world' property of high clustering combined with short path lengths [1]. The most widely used models of small world networks are the Watts-Strogatz (WS) model [2] and a slight variant of the WS model known as the Newman-Watts (NW) model [3]. In this brief paper we describe an alternative, but essentially equivalent, formulation of the WS model in which the presence or absence of connections is determined independently for each possible edge according to a distance-dependent probability of connection. This simplifies the generation of both directed and undirected small-world networks, in the same way that the $G(n, p)$ model of random 'Erdős-Rényi' (ER) networks with $n$ nodes and a fixed probability of connection $p$ [4] is often simpler to analyze than the $G(n, m)$ ER model with fixed number of edges $m$ [5] because in the former the edges are completely independent of one another. The reformulated WS model is mathematically 'cleaner' than existing formulations and there are clear advantages to replacing current implementations of the WS model. However, constructing undirected small-world networks is already straightforward and the concept of a small world loses its significance in densely connected networks (Fig. 1). Therefore, the primary value of the reformulated WS model is in highlighting an essential feature of the WS model that has previously been overlooked, and making it possible to derive exact expressions for quantities such as the degree and motif distributions and global clustering coefficient for both directed and undirected networks in terms of model parameters.

Consider a network with $L$ nodes labeled $i = 0, \ldots, L-1$. In the usual WS model of an undirected small-world network [2] the nodes of the network are placed on a ring lattice with periodic boundary conditions (i.e., node $L$ is identified with node 0) and each node is initially connected to its $K$ (conveniently taken to be even) nearest neighbors on the lattice. This is the regular lattice limit of the WS model. Next, each edge $u \leftrightarrow v$ is 'rewired' with probability $\beta$ to $u \leftrightarrow w$ where $w$ can be any node $w \neq u, v$ as long as the connection $u \leftrightarrow w$ does not already exist (it is also common to replace both $u$ and $v$, and the restriction on multiple edges between the same pair of nodes is not always enforced [6]). Since $w$, unlike $v$, can be located anywhere on the lattice and is not necessarily one of the $K$ nearest neighbors of $u$, the rewired connection $u \leftrightarrow w$ often acts as a shortcut within the network. By varying the parameter $\beta$ from 0 to 1, it becomes possible to interpolate between the original regular lattice limit ($\beta = 0$) and the completely random ER limit ($\beta = 1$). Watts and Strogatz found [2] that there is a relatively large range for the value of $\beta$ over which the average path length is short and clustering is high (Fig. 1a), so that the network is said to possess the 'small-world' property.

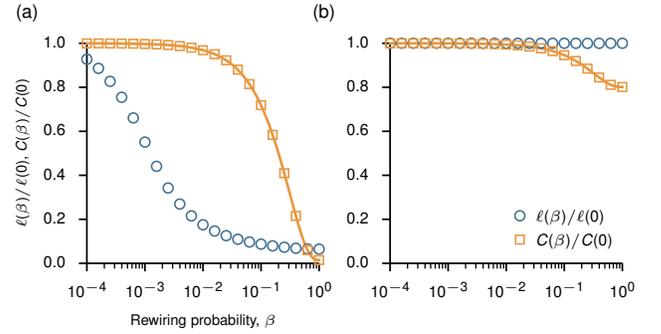

FIG. 1: (Color online) Average path length $\ell(\beta)$ (circles) and global clustering coefficient $C(\beta)$ (squares) in the reformulated, undirected Watts-Strogatz model described by Eq. (1), for edge density (a) $p_0 = 0.01$ and (b) $p_0 = 0.6$. The corresponding directed network behaves identically with respect to average path length and global clustering coefficient. The average path length and global clustering coefficient are both normalized by their values in the regular lattice limit, $\ell(0)$ and $C(0)$, respectively. For easy comparison to the original model in Ref. 2 we have chosen the network size to be $L = 1001$, so that $K = 10$ in (a). The exact global clustering coefficients given by Eq. (9) are shown as solid lines. Note that for $p_0 \geq 0.5$ the average path length is essentially $\ell(\beta) = 2 - p_0$ regardless of the rewiring probability $\beta$, while the global clustering coefficient takes values in a relatively small range. Thus the notion of a 'small world' loses its significance in densely connected networks like the one shown in (b).

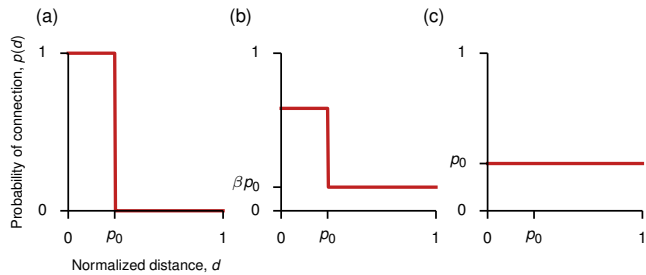

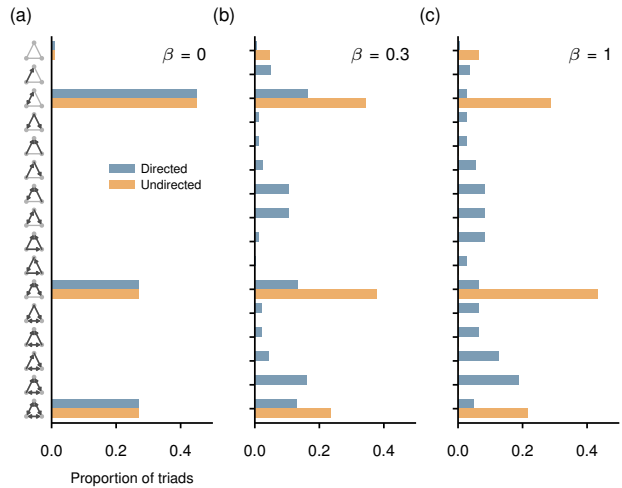

FIG. 2: (Color online) Distance-dependent formulation of the Watts-Strogatz (WS) model. (a) Regular lattice limit, $\beta = 0$. (b) Intermediate $\beta$, here 0.5. (c) Erdős-Rényi limit, $\beta = 1$. In all cases $p_0 = 0.3$ for illustration purposes; in reality, most discussions of the WS model deal with situations where $p_0 \ll 1$.

In the NW variant of the WS model [3], the only difference is that, instead of rewiring the edges of the regular lattice, the shortcuts are superposed on the original regular lattice. This simplifies certain analytical calculations by ensuring that the network always remains connected after rewiring, i.e., there is always a path through the network that connects two nodes. On the other hand, the (typically negligible) cost of this simplification is that there is no true random limit, even at $\beta = 1$, because the original regular lattice always remains. Indeed, it is the case in many existing implementations of the WS model that the true ER limit does not exist [6].

Although the idea of rewiring edges makes the WS model very intuitive, in practice it introduces some inconvenient features into the model that only become apparent when the network is small and the connectivity dense. These effects are almost always ignored because most networks that have been studied are large and sparse (with the notable exception of the areal network of the mammalian cerebral cortex [7], which has $L \sim 100$ and $p_0 \sim 0.6$), but are nevertheless undesirable. To illustrate this point, consider the random limit of the WS model reached by rewiring all edges, $\beta = 1$. There are two issues: first, from a practical point of view the process of rewiring requires checking existing connections to ensure that there are no multiple edges (one can also simply accept multiple edges), and second, from a mathematical point of view the edges are not independent because the network is constrained to approach the $G(L, LK/2)$ ER network with a fixed number of edges (inherited from the regular lattice), rather than the $G(L, p_0)$ network with a fixed probability of connection where no such global constraint exists.

FIG. 3: (Color online) Example of the exact triad distribution computed using Eq. (3) (directed, upper bars) and Eq. (4) (undirected, lower bars). In both cases the network size is $L = 1001$ and we illustrate the distribution for the densely connected case of $p_0 = 0.6$ as in Fig. 1b. (a) Regular lattice limit, $\beta = 0$. (b) Intermediate regime, $\beta = 0.3$. (c) Random ER limit, $\beta = 1$.

An alternative, but essentially equivalent, way to understand the WS model is to consider each edge to be present with a probability that depends, in a simple way, on the distance between the nodes. Here, the distance $D$ between two nodes is the shortest number of steps it takes to get from one node to another along the ring, $D_{ij} = \min(|i-j|, L-|i-j|)$. It is convenient to normalize the distance by $D_{\max} = \lfloor L/2 \rfloor$, so that $d_{ij} = D_{ij}/D_{\max}$ with $0 \leq d_{ij} \leq 1$. Let $p_0 = K/(L-1)$. Then the probability that an edge exists between two nodes is given by

$$p_{ij} = p(d_{ij}) = \beta p_0 + (1-\beta)\Theta(p_0 - d_{ij}), \quad (1)$$

where $\Theta(x)$ is the Heaviside step function with $\Theta(x) = 1$ if $x \geq 0$ and zero otherwise. The fact that each edge is chosen independently and is obviously either 0 or 1 makes this interpretation of the WS model mathematically more appealing than the rewiring formulation. In the regular lattice limit $\beta = 0$ and $p_{\text{RL}}(d) = \Theta(p_0 - d)$, i.e., only nodes within a distance $p_0$ are connected, while in the ER limit $\beta = 1$ and $p_{\text{ER}}(d) = p_0$. For intermediate values of $\beta$, the probability of two nodes separated by a distance greater than $p_0$ being connected is $\beta p_0$ while the probability of two nodes within a radius $p_0$ being connected is the sum of the original lattice minus rewired contribution, $1 - \beta$, and the rewired contribution, $\beta p_0$. It can be checked that $\int_0^1 dx\, p(x) = p_0$.

Eq. (1) can be used for both directed and undirected networks: for directed networks the edges $u \to v$ and $u \leftarrow v$ are determined independently, while for undirected networks both edges are determined together. This is a more natural, and consistent, way to generate directed networks compared to the rewiring algorithm. Moreover, Eq. (1) is not restricted to one dimension, and generalizes to higher dimensions with appropriate modifications. Interestingly, the $\beta = 0$ limit of Eq. (1) is the rule used for both random geometric graphs [8] and scale-free networks constructed in the framework of hyperbolic geometry [9], and it is of interest to investigate the properties of spatially embedded networks generated according to Eq. (1).

The three possible regimes for $\beta$ are illustrated in Fig. 2, which makes clear the 'geometry' of the WS model. In all cases, the probability of two nodes being connected by an edge is a step function of the distance between the nodes. In the random limit $\beta = 1$, however, the two 'steps' have equal probability. Although not surprising, Fig. 2 illustrates the essential lack of spatial realism in WS-type small-world networks; notably, the probability of connection depends only on a cutoff radius (defined by $K$ or $p_0$) and the probability of connection does not go to 1 as $d \to 0$. According to this picture, it is most natural to allow self-connections to be chosen with probability $p(0)$ (in which case $p_0 = K/L$), but this may not be the correct choice in all situations. In particular, it must be emphasized that distances in WS-type models do not necessarily represent physical distances.

Because every connection is chosen independently, the formulation in terms of distances allows us to easily average over ensembles and derive exact expressions for several quantities of interest by inspection. For instance, the exact degree (either out-degree or in-degree if the network is directed) distribution $f(k)$ is clearly

$$f(k) = \sum_{k_1=0}^{k} \binom{K}{k_1} p_1^{k_1}(1-p_1)^{K-k_1} \binom{L'-K}{k-k_1} p_2^{k-k_1}(1-p_2)^{L'-K-(k-k_1)}, \qquad 0 \le k \le L', \qquad (2)$$

where $p_1 = \beta p_0 + 1 - \beta$, $p_2 = \beta p_0$, and $L' = L$ if self-connections are counted and $L' = L - 1$ otherwise. It is understood in Eq. (2) that the binomial coefficient $\binom{a}{b} = 0$ if $b > a$. Eq. (2) is similar in form to the well-known result from Ref. 10. As a check, in the regular lattice limit $\beta \to 0$ Eq. (2) becomes $f(k) = 1$ if $k = K$ and 0 otherwise, while in the ER limit $\beta = 1$ Eq. (2) reduces to the binomial distribution $f(k) = \binom{L'}{k} p_0^k (1-p_0)^{L'-k}$.

A more interesting benefit of the reformulated WS model is that it becomes straightforward to express the motif distribution by generalizing the motif distribution for the ER random network. In the directed case the proportion of triads (three-node combinations) with a given motif is given by

$$P_{\rm d}(t) = \frac{m}{(L-1)(L-2)/2} \sum_{0<j<k} p_{0j}^{n_1}(1-p_{0j})^{2-n_1} p_{0k}^{n_2}(1-p_{0k})^{2-n_2} p_{jk}^{n_3}(1-p_{jk})^{2-n_3}, \qquad (3)$$

where the combinatorial factor $m$ and number of edges $n = (n_1, n_2, n_3)$ for each triad $t = (m, n)$ are given in Table I, and we have dropped the subscript $t$ for notational simplicity in Eq. (3). We have used translational invariance and reflection symmetry to simplify the sum. In the ER limit, $\beta = 1$, Eq. (3) reduces to $m p_0^{n_1+n_2+n_3}(1-p_0)^{6-(n_1+n_2+n_3)}$. Similarly, for the far less interesting case (because there are fewer possible motifs) of undirected networks the occurrence of triads is given by

$$P_{\rm ud}(t) = \frac{m}{(L-1)(L-2)/2} \sum_{0<j<k} p_{0j}^{n_1}(1-p_{0j})^{1-n_1} p_{0k}^{n_2}(1-p_{0k})^{1-n_2} p_{jk}^{n_3}(1-p_{jk})^{1-n_3}. \qquad (4)$$

Again the appropriate parameters are given in Table I, and in the ER limit Eq. (4) reduces to $m p_0^{n_1+n_2+n_3}(1-p_0)^{3-(n_1+n_2+n_3)}$. In this way, moreover, the occurrence of motifs of any size can be calculated if the combinatorial factors are known. An example of computing the exact triad distribution using Eqs. (3) and (4) is presented in Fig. 3.

| Triad $t$ | name | Directed | | Undirected | |
|---|---|---|---|---|---|
| | | $m_t$ | $n_t$ | $m_t$ | $n_t$ |
|  | 003 | 1 | $(0,0,0)$ | 1 | $(0,0,0)$ |
|  | 012 | 6 | $(1,0,0)$ | - | - |
|  | 102 | 3 | $(2,0,0)$ | 3 | $(1,0,0)$ |
|  | 021D | 3 | $(1,1,0)$ | - | - |
|  | 021U | 3 | $(1,1,0)$ | - | - |
|  | 021C | 6 | $(1,1,0)$ | - | - |
|  | 111D | 6 | $(2,1,0)$ | - | - |
|  | 111U | 6 | $(2,1,0)$ | - | - |
|  | 030T | 6 | $(1,1,1)$ | - | - |
|  | 030C | 2 | $(1,1,1)$ | - | - |
|  | 201 | 3 | $(2,2,0)$ | 3 | $(1,1,0)$ |
|  | 120D | 3 | $(2,1,1)$ | - | - |
|  | 120U | 3 | $(2,1,1)$ | - | - |
|  | 120C | 6 | $(2,1,1)$ | - | - |
|  | 210 | 6 | $(2,2,1)$ | - | - |
|  | 300 | 1 | $(2,2,2)$ | 1 | $(1,1,1)$ |

TABLE I: Parameters for computing the triad distribution. The commonly used designation for each triad [11, 12], which indicates the number of double, single, and zero edges, is also given for convenience. $m_t$ and $n_t$ are the combinatorial multiplicity and number of edges, respectively.

The global clustering coefficient can be calculated in a similar manner to the triad distribution. There are several different definitions for the global clustering coefficient of a network that differ slightly in their detail, but they all reflect the probability that two nodes $u$ and $v$ are connected when $u$ and $v$ are both connected to a third node $w$. Alternatively, the global clustering coefficient measures the ratio of the number of closed triplets to the number of connected triplets. Here we consider the definition from Ref. 13, which is based on a commonly used definition of local clustering coefficient described in Ref. 14 and has several desirable properties, including applicability to both directed and undirected networks and the fact that its value is $p_0$ in the random ER limit of both directed and undirected cases. This is in contrast to several calculations in which this does not hold, e.g., in Ref. 6.

Let the network adjacency matrix be $A$, with transpose $A^T$. Then the global clustering coefficient can be written as [13, 14]

$$C(A) = \frac{\frac{1}{2}\sum_i [(A+A^T)^3]_{ii}}{\sum_i [d_i^{\text{tot}}(d_i^{\text{tot}}-1) - 2d_i^{\leftrightarrow}]}, \quad (5)$$

where $d_i^{\text{tot}} = \sum_j (A+A^T)_{ij}$ is the total degree (sum of in-degree and out-degree) of node $i$ and $d_i^{\leftrightarrow} = (A^2)_{ii}$ is the number of bilateral edges from node $i$. We will express the clustering coefficient in terms of the triad distribution with adjusted combinatorial factors, reflecting the fact, for example, that in the numerator of Eq. (5) bilateral edges contribute twice to the sum. In the directed case

$$C_d = \frac{S_{\text{closed}}}{S_{\text{connected}}}, \quad (6)$$

where

$$S_{\text{closed}} = \frac{1}{2}\Big[P_d(030T) + P_d(030C) + 2P_d(120D) + 2P_d(120U) + 2P_d(120C) + 4P_d(210) + 8P_d(300)\Big], \quad (7)$$

$$S_{\text{connected}} = \frac{1}{3}\Big[P_d(021D) + P_d(021U) + P_d(021C) + 2P_d(111D) + 2P_d(111U) + 3P_d(030T) + 3P_d(030C)$$
$$+ 4P_d(201) + 5P_d(120D) + 5P_d(120U) + 5P_d(120C) + 8P_d(210) + 12P_d(300)\Big]. \quad (8)$$

In the undirected case we have the simpler expression

$$C_{\text{ud}} = \frac{3P_{\text{ud}}(300)}{3P_{\text{ud}}(300) + P_{\text{ud}}(201)}, \quad (9)$$

which corresponds to the classical definition of transitivity for undirected networks [1, 6]. In the regular lattice limit, $\beta = 0$, Eq. (9) reduces to the usual value [10] $C = 3(1-\delta)/4$, $\delta = 1/(K-1)$ if $p_0 \le 2/3$. See Fig. 1 for an example of the application of Eq. (9). We note that Eqs. (6-8) and Eq. (9) are general expressions for the global clustering coefficient in terms of the triad distribution, which to our knowledge have not been reported previously. Moreover, although we have given separate expressions for the directed and undirected cases $C_d$ and

$C_{ud}$ as functions of the parameters $L$, $p_0$, and $\beta$ are identical in our model.

In principle, a similar approach can be used to calculate the occurrence of shortest paths of length 1, 2, . . . . For instance, the probability that a pair of nodes is connected by a path of length 1 is simply $q_1 = p_0$, and the probability that a pair of nodes is connected by a path of length 2 (but not by a path of length 1) is

$$q_2 = \frac{1}{L-1} \sum_{j \neq 0} (1-p_{0j}) \left[ 1 - \prod_{k \neq 0, j} (1 - p_{0k} p_{kj}) \right]. \quad (10)$$

Thus in the ER limit $q_2(\beta=1) = (1-p_0)[1-(1-p_0^2)^{L-2}]$. It is of interest to use this line of reasoning to express the exact average path length in a tractable manner. Note that we can perform (again, in principle) this calculation conditioned on the network being connected.

Finally, there may be situations where it is convenient to rewrite Eq. (1) in a way that emphasizes the periodic nature of the ring lattice. We can obtain the same networks as those generated by Eq. (1) if we 'transform coordinates' to

$$z_{ij} = \sin^2\left(\frac{i-j}{L}\pi\right), \qquad \mu_0 = \sin^2\left(\frac{p_0}{2}\pi\right). \quad (11)$$

Then the probability of connection between nodes $i$ and $j$ is given by

$$p_{ij} = \beta p_0 + (1-\beta)\Theta(\mu_0 - z_{ij}). \quad (12)$$

In conclusion, we have introduced an alternative formulation of the widely used Watts-Strogatz model for small-world networks. Although much of the material presented in this work is quite straightforward, in our view this formulation highlights an essential feature of the WS model that has previously been overlooked, and has the practical benefit of simplifying the generation of both directed and undirected WS-type small-world networks. We have also shown that reformulating the WS model in terms of a distance-dependent probability of connection so that each edge is present or absent independently of the others makes it straightforward to derive exact expressions for quantities such as the degree and motif distributions and global clustering coefficient for both directed and undirected networks. We believe that in many settings where the rewiring algorithm is currently used, there are advantages to using this mathematically cleaner interpretation of the WS model.

This work was supported by ONR grant N00014-13-1-0297, NIH grant MH062349, and the Swartz Foundation. We thank M. Javarone for calling our attention to Ref. 9 and B. Sonnenschein for clarification on the two types of ER networks.